\documentclass[conference]{IEEEtran}
\usepackage{cite}
\usepackage{amsmath,amssymb,amsfonts}
\usepackage{algorithmic}
\usepackage{graphicx}
\usepackage{textcomp}
\usepackage{xcolor}
\usepackage{amsfonts}
\usepackage{url}
\usepackage{hyperref}
\usepackage{booktabs}
\usepackage{float}
\usepackage{amsmath}
\usepackage{array}
\usepackage{algorithm}

\def\BibTeX{{\rm B\kern-.05em{\sc i\kern-.025em b}\kern-.08em
    T\kern-.1667em\lower.7ex\hbox{E}\kern-.125emX}}
\begin{document}

\makeatletter
\newcommand{\linebreakand}{%
  \end{@IEEEauthorhalign}
  \hfill\mbox{}\par
  \mbox{}\hfill\begin{@IEEEauthorhalign}
}
\makeatother

\title{SecureFixAgent: A Hybrid LLM Agent for Automated Python Static Vulnerability Repair}

\author{\IEEEauthorblockN{1\textsuperscript{st} Jugal Gajjar}
\IEEEauthorblockA{\textit{Computer Science Department} \\
\textit{The George Washington University}\\
Washington D.C, USA \\
jugal.gajjar@gwu.edu}
\and
\IEEEauthorblockN{2\textsuperscript{nd} Kamalasankari Subramaniakuppusamy}
\IEEEauthorblockA{\textit{Computer Science Department} \\
\textit{The George Washington University}\\
Washington D.C, USA \\
kamalasankaris@gwu.edu}
\and
\IEEEauthorblockN{3\textsuperscript{rd} Relsy Puthal}
\IEEEauthorblockA{\textit{Applied Economics Department} \\
\textit{The George Washington University}\\
Washington D.C, USA \\
relsy.puthal@gwu.edu}
\linebreakand
\IEEEauthorblockN{4\textsuperscript{th} Kaustik Ranaware}
\IEEEauthorblockA{\textit{Computer Science Department} \\
\textit{The George Washington University}\\
Washington D.C, USA \\
k.ranaware@gwu.edu}
}

\maketitle

\begin{abstract}
Modern software development pipelines face growing challenges in securing large codebases with extensive dependencies. Static analysis tools like Bandit are effective at vulnerability detection but suffer from high false positives and lack repair capabilities. Large Language Models (LLMs), in contrast, can suggest fixes but often hallucinate changes and lack self-validation. We present SecureFixAgent, a hybrid repair framework integrating Bandit with lightweight local LLMs ($<$8B parameters) in an iterative detect--repair--validate loop. To improve precision, we apply parameter-efficient LoRA-based fine-tuning on a diverse, curated dataset spanning multiple Python project domains, mitigating dataset bias and reducing unnecessary edits. SecureFixAgent uses Bandit for detection, the LLM for candidate fixes with explanations, and Bandit re-validation for verification, all executed locally to preserve privacy and reduce cloud reliance. Experiments show SecureFixAgent reduces false positives by 10.8\% over static analysis, improves fix accuracy by 13.51\%, and lowers false positives by 5.46\% compared to pre-trained LLMs, typically converging within three iterations. Beyond metrics, developer studies rate explanation quality 4.5/5, highlighting its value for human trust and adoption. By combining verifiable security improvements with transparent rationale in a resource-efficient local framework, SecureFixAgent advances trustworthy, automated vulnerability remediation for modern pipelines.
\end{abstract}

\begin{IEEEkeywords}
static program analysis, large language models, automated code repair, LLM agents, software vulnerability detection
\end{IEEEkeywords}

\section{Introduction}
\label{intro}
The rapid expansion of modern software systems has resulted in increasingly large and complex codebases, often composed of numerous third-party dependencies \cite{he2023automating}. This growth inevitably widens the potential vulnerability surface, making security assurance a critical component of the software development lifecycle. Vulnerabilities that escape early detection can lead to severe consequences in production environments, including data breaches, financial losses, and reputational damage \cite{gratton2025cybercrime}. As a result, organizations are adopting automated tools to identify and mitigate security risks before deployment.

Static analysis tools, such as Bandit \cite{bandit} for Python, are widely used for early-stage vulnerability detection. These tools operate by scanning source code against predefined security rules and patterns, offering high precision in identifying certain classes of vulnerabilities. However, static analyzers suffer from notable drawbacks: they tend to generate high false-positive rates, burdening developers with unnecessary manual triage, and they lack inherent capabilities for automated repair \cite{harzevili2023}. Consequently, while they are effective at highlighting potential issues, the remediation process remains manual, time-consuming, and prone to human oversight.

Recent advances in Large Language Models (LLMs) have shown promise in automating code repair. Code-specialized LLMs can analyze vulnerable snippets, reason about flaws, and generate candidate fixes \cite{zhou2025large}. Industry tools such as GitHub Copilot have begun experimenting with vulnerability-aware suggestions and automated repairs \cite{github2024security}, reflecting the growing commercial interest in this domain. However, LLM-based approaches face two major limitations in security-critical contexts. First, they frequently produce hallucinated fixes—syntactically valid but semantically ineffective changes that fail to eliminate the underlying vulnerability or introduce new ones \cite{chen2024deep}. Second, they lack built-in self-verification mechanisms, resulting in unvalidated patches that require human review.

To address these limitations, we introduce SecureFixAgent, a hybrid framework that merges the strengths of static analysis and LLMs for automated code repair through a detect--repair--validate loop. SecureFixAgent leverages Bandit \cite{bandit} to perform initial vulnerability detection, utilizes a locally deployed lightweight code LLM ($<$8B parameters) to propose context-aware fixes accompanied by human-readable explanations, and then re-applies Bandit \cite{bandit} to validate the correctness of the patch. If the vulnerability persists, the process iterates with refined prompts and contextual feedback until a secure fix is reliably achieved. All inference is performed locally, ensuring privacy preservation and eliminating reliance on external cloud-based APIs.

The key contributions of this work are threefold. First, we design and implement a hybrid vulnerability repair pipeline that autonomously detects and repairs Python code vulnerabilities while maintaining human-readable, explainable output. Second, we integrate multiple local code LLMs with Bandit to create a validation-driven repair mechanism that minimizes hallucinated fixes. Third, we present an empirical evaluation on the curated vulnerable code samples, demonstrating that SecureFixAgent improves fix accuracy and reduces false positives compared to either static analysis or LLM-based approaches alone.

\section{Related Work}
\label{rel_work}
Traditional approaches to vulnerability detection have relied on static, rule-based analysis. Tools such as Bandit \cite{bandit} and PyLint \cite{pylint} scan Python code for insecure patterns (e.g., hardcoded credentials, insecure deserialization). While precise for well-defined vulnerability classes, these systems often yield high false-positive rates and provide no automated repair \cite{harzevili2023}.

Dynamic analysis complements static methods by executing code in controlled environments (e.g., OWASP ZAP \cite{zap}, Burp Suite \cite{burpsuite}), enabling detection of runtime issues such as race conditions and memory leaks that static scans may miss. However, adoption remains limited due to brittle execution setups, performance overhead, and incomplete code coverage \cite{alashjaee2019}. This leads to coverage gaps in real-world deployments, especially for large, distributed systems.

Machine learning and deep learning approaches broadened flexibility by learning vulnerability patterns from datasets such as Juliet \cite{juliet} and Draper VDISC \cite{draper}. VulDeePecker \cite{vuldeepecker} demonstrated semantic-level flaw detection using code gadget representations, achieving higher precision, though at the cost of requiring extensive labeled datasets and offering no automated remediation.

Transformer-based models (CodeBERT \cite{codebert}, CodeT5 \cite{codet5}, and PolyCoder \cite{polycoder}) advanced code intelligence tasks, while security-focused variants such as VulBERTa \cite{vulberta} and MalBERT \cite{malbert} targeted vulnerability detection. APR systems such as T5APR \cite{t5apr} and Repairnator \cite{repairnator} generate patches, and commercial tools like GitHub Copilot \cite{githubcopilot} now incorporate vulnerability-aware suggestions. However, these systems still face three persistent gaps: (1) lack of rigorous, security-specific validation of generated patches, often leading to unverified or incomplete fixes \cite{chen2024deep}; (2) reliance on large, cloud-hosted models, introducing privacy, cost, and latency concerns \cite{yao2024survey}; and (3) limited mechanisms for iterative refinement when initial repairs fail \cite{malcodeai}.

Recent research has begun to couple vulnerability detection with semantic reasoning and natural language explanations. For example, MalCodeAI \cite{malcodeai} decomposes code into functional units, applies LLM-based detectors, and produces both patches and human-readable rationales, improving transparency. Other work integrates user feedback into LLM-assisted vulnerability remediation processes \cite{wang2025llm}, aiming to boost repair satisfaction and trustworthiness. However, these systems still operate primarily in a \textit{single-pass repair mode}, lacking the iterative validation necessary to reliably handle persistent or complex vulnerabilities in security-critical code.

In contrast, SecureFixAgent explicitly addresses these limitations by combining static analysis feedback with iterative LLM-driven repair and validation, all executed locally to preserve privacy. Unlike Copilot or Repairnator, which either lack validation or rely on cloud-scale models, SecureFixAgent is designed for on-premise feasibility, explainability, and resource efficiency. To make this distinction clearer, Table~\ref{tab:comparison} summarizes how SecureFixAgent compares with representative prior systems across validation, deployment, and refinement dimensions.

\begin{table}[!t]
\caption{Comparison of SecureFixAgent with Prior Systems}
\centering
\begin{tabular}{|l|c|c|c|}
\hline
System & Validation & Deployment & Iterative Repair \\
\hline
GitHub Copilot & $\times$ & Cloud & $\times$ \\
Repairnator & Partial & Server & $\times$ \\
MalCodeAI & $\times$ & Local LLM & $\times$ \\
SecureFixAgent & $\checkmark$ Bandit & Local LLM & $\checkmark$ Loop \\
\hline
\end{tabular}
\label{tab:comparison}
\end{table}

\section{Methodology}
\label{method}
We present SecureFixAgent, a hybrid vulnerability detection and repair framework that integrates static analysis with open-source, local, and resource-efficient LLMs in an iterative detect--repair--validate loop for robustness. SecureFixAgent executes fully on-premise, ensuring privacy compliance and low latency with models capped at 8B parameters. Quantization methods (INT8, 4-bit, mixed precision) further reduce memory and compute costs with minimal performance loss, enabling deployment on consumer-grade hardware. Unlike monolithic LLM-based patching systems, our approach enforces detection and patch correctness through repeated static revalidation, converging iteratively toward verified vulnerability resolution with accompanying human-readable rationales. Figure \ref{fig:architecture} illustrates the architecture of our proposed framework and data flow among its core components.

\begin{figure*}[!t]
\centering
\includegraphics[width=0.9\textwidth, height=2.4in]{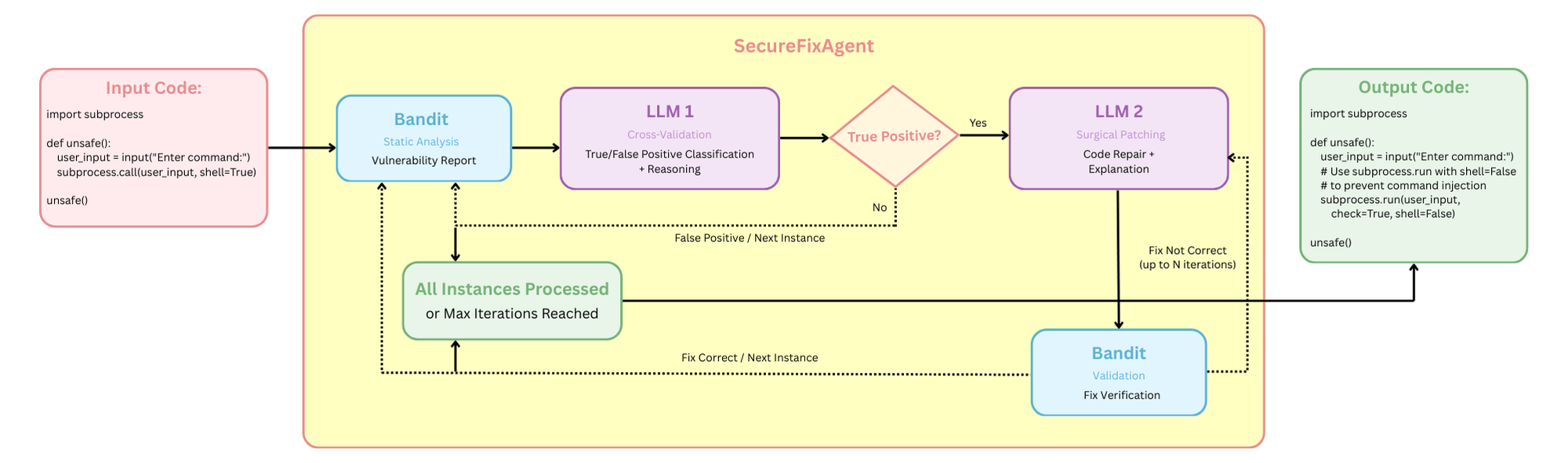}
\caption{Architecture of \emph{SecureFixAgent}, integrating static analysis with locally hosted LLMs in an iterative detect--repair--validate loop. Bandit identifies candidate vulnerabilities, LLM~1 cross-validates reports, and LLM~2 performs targeted patching with explanations. Patched code is re-verified by Bandit until convergence or a maximum iteration limit is reached, yielding traceable and verified repairs.}
\label{fig:architecture}
\end{figure*}

\subsection{System Architecture}

The SecureFixAgent pipeline consists of four tightly integrated stages. Initially, Bandit statically analyzes Python source code to identify potential vulnerabilities, generating a structured report. Next, a locally hosted, code-specialized LLM instance interprets the report to validate the detection and to synthesize minimal, targeted patches accompanied by concise, human-readable explanations. The validation stage re-scans the patched code using Bandit to confirm vulnerability resolution. If vulnerabilities persist, the system iterates with updated context until convergence is reached or a predefined iteration limit is met. The final output stage produces the fully patched code alongside a concise technical explanation.

\subsection{Pipeline Flow}

Starting from an input Python file, Bandit performs vulnerability scanning to produce a detailed report. The report and original source are provided to the LLM, which proposes minimal corrective changes with explanatory comments for the truly positive detections. The patched code is then re-analyzed by Bandit; if unresolved vulnerabilities remain, this detect--repair--validate loop repeats. Iteration continues until all issues are fixed or the maximum iteration count is reached, balancing fix thoroughness with computational efficiency.

\subsection{Algorithm Specification}
Algorithm \ref{alg:securefixagent} formalizes the iterative detect--repair--validate process described in the Pipeline Flow. Unlike monolithic repair strategies that attempt to patch all detected vulnerabilities in a single pass, this variant processes each vulnerability individually. For every element flagged by Bandit, the original code segment and the relevant portion of the Bandit report are provided to the LLM, which generates a patch targeted exclusively at that instance. This per-vulnerability approach enables finer-grained control, minimizes unintended code modifications, and supports precise tracking of which changes resolve specific vulnerabilities. The output package generated after convergence contains (i) the original code, (ii) the fully patched code, (iii) the corresponding Bandit report for each iteration, and (iv) LLM explanations for each remediation, enabling full traceability and auditability of the repair process.

\begin{algorithm}[ht]
\caption{SecureFixAgent: Iterative Fine-Grained Vulnerability Detection and Repair}
\label{alg:securefixagent}
\begin{algorithmic}[1]
\renewcommand{\algorithmicrequire}{\textbf{Input:}}
\renewcommand{\algorithmicensure}{\textbf{Output:}}
\REQUIRE Python source code file $C$, maximum iteration limit $N$
\ENSURE Original code $C_{orig}$, final patched code $C$, iteration-wise Bandit reports, LLM-generated explanations
\\ \textit{Initialisation} :
\STATE $i \gets 0$
\STATE $C_{orig} \gets C$
\\ \textit{Iterative Detection and Repair Loop} :
\REPEAT
    \STATE Run Bandit on $C$ to produce vulnerability report $R$
    \IF{$R$ is empty}
        \STATE \textbf{break} \COMMENT{No vulnerabilities detected}
    \ENDIF
    \FOR{each vulnerability $v \in R$}
        \STATE Extract vulnerable code segment $S_v$ from $C$
        \STATE Retrieve corresponding report excerpt $R_v$
        \STATE Provide $S_v$, $R_v$, and $C$ to locally hosted LLM
        \STATE LLM cross-validates the report excerpt $R_v$ with human-readable explanation
        \IF{$R_v$ is True Positive}
            \STATE LLM generates patched segment $S'_v$ with human-readable explanation
            \STATE Replace $S_v$ in $C$ with $S'_v$
        \ENDIF
    \ENDFOR
    \STATE Save Bandit report $R$ for iteration $i$
    \STATE $i \gets i + 1$
\UNTIL{$R$ is empty \textbf{or} $i \geq N$}
\\ \textit{Finalization} :
\STATE Run final Bandit scan to confirm all fixes
\STATE Generate output package containing original code $C_{orig}$, final code $C$, all Bandit reports, and LLM explanations
\RETURN Final output package
\end{algorithmic}
\end{algorithm}

\subsection{LLM Configuration and Fine-Tuning}

We evaluate several locally executable, code-specialized LLMs—including DeepSeek Coder \cite{deepseekcoder}, Qwen2.5-Coder \cite{qwen25}, CodeLlama \cite{codellama}, and CodeGemma \cite{codegemma}—with parameter sizes capped at 8 billion to ensure feasible deployment on standard workstations. In addition to testing base (pre-trained) model performance, we perform supervised fine-tuning using a curated dataset of Python vulnerabilities and corresponding fixes, sourced from both synthetic injections and real-world CVE patches. Fine-tuning is implemented using the Apple MLX framework \cite{mlx} for M-series optimization, combined with Low-Rank Adaptation (LoRA) \cite{lora} to minimize computational overhead while preserving model performance. This approach improves the models’ ability to generate precise, minimal, and contextually appropriate repairs while adhering to security best practices. All model inference, whether from base or fine-tuned variants, is executed entirely on-device, eliminating reliance on cloud services and mitigating sensitive data exposure.

\subsection{Prompt Design}

To minimize hallucinations and promote reproducibility, prompts are carefully structured to guide the LLM through three core tasks: interpreting individual Bandit-reported vulnerabilities, proposing minimal code changes targeted to resolve each issue without altering intended functionality, and generating concise, human-readable explanations. Outputs are explicitly requested in structured formats—e.g., JSON-like dictionaries embedded in valid Python code—enabling programmatic parsing and validation of LLM responses. This structured output design is essential for automated downstream processing and verification.

\subsection{Privacy and Efficiency}

By confining all inference to local hardware, SecureFixAgent adheres to stringent code confidentiality requirements, preventing potential leakage of proprietary source code. The iterative pipeline is optimized for low latency per cycle, supporting integration into continuous integration and deployment (CI/CD) workflows where timely automated remediation is essential. All intermediate artifacts are encrypted at rest and in memory with AES-128 for reliable, efficient protection, and no telemetry is transmitted externally, preserving organizational privacy.

\section{Experiments}
\label{experiments}
The experimental evaluation of SecureFixAgent was conducted in two hardware environments. The first consisted of an Apple M-series system, offering high-efficiency ARM-based processing for local inference. The second was a CUDA-enabled GPU workstation equipped with an NVIDIA GPU, enabling accelerated inference for larger models and batch processing. These setups ensured feasibility within on-premise constraints while supporting evaluation of models up to 8 billion parameters. We ensured runtime efficiency while keeping memory usage viable: average per-iteration latency remained practical for CI/CD deployment, with peak memory during model execution ranging from 6 to 14 GB for the best-performing models and up to 40 GB during fine-tuning.

The dataset was drawn from two primary sources. The first source was a synthetically generated corpus in which vulnerability injection locations and types were randomized across modules and systematically injected into Python programs to simulate realistic distributions. This allowed controlled experimentation and reproducibility by ensuring that vulnerability locations and types were known a priori. The second source consisted of real-world Python repositories, including CVE-Bench \cite{cvebench}, PySecDB \cite{sun2023exploring}, and SecurityEval \cite{siddiq2022seceval}, containing publicly disclosed Common Vulnerabilities and Exposures (CVEs), providing realistic and diverse vulnerability patterns for testing the efficacy of our proposed system.

\begin{table}[htbp]
\caption{Data Summary}
\begin{center}
\begin{tabular}{|l|l|l|}
\hline
\textbf{Data Source} & \textbf{\# Samples} & \textbf{\# Vulnerability Types} \\
\hline
Synthetic (Injected) & 740 & 52 \\
CVE-Bench & 40 & 8 \\
PySecDB & 1258 & 119 \\
SecurityEval & 130 & 75 \\
\hline
\end{tabular}
\label{tab:data_summary}
\end{center}
\end{table}

SecureFixAgent was evaluated using multiple open-source, code-specialized LLMs: DeepSeek Coder (1.3B, 6.7B) \cite{deepseekcoder}, Qwen 2.5 Coder (3B, 7B) \cite{qwen25}, CodeLlama (7B) \cite{codellama}, and CodeGemma (7B) \cite{codegemma}. Each model was tested in its raw, instruction-tuned form to establish baseline performance. Subsequently, models were fine-tuned with Apple MLX \cite{mlx} using LoRA \cite{lora}, leveraging the combined synthetic and real-world vulnerability–repair dataset to improve repair accuracy and reduce hallucinated changes.

Three system configurations were compared: (1) Bandit alone for vulnerability detection without automated repair; (2) LLM-based repair without static validation (single-pass patch generation); and (3) the full SecureFixAgent pipeline integrating Bandit detection with iterative, validation-driven LLM repair. Performance was assessed using four primary metrics—fix accuracy, false-positive rate, iterations to convergence, and explanation clarity (developer-rated Likert scale)—with statistical significance evaluated via paired $t$-tests ($p<0.05$). Additionally, while commercial tools such as GitHub Copilot were not directly benchmarked due to closed-source constraints, qualitative comparisons were discussed in terms of privacy, iterative validation, and repair explainability.

Results were reported for both raw and fine-tuned models, enabling clear evaluation of the impact of domain adaptation on SecureFixAgent’s repair effectiveness.

\section{Results}
\label{results}
The experimental results demonstrate that integrating static analysis with iterative, locally executed LLM-based repair substantially improves vulnerability resolution compared to either approach alone. Across all evaluated models, SecureFixAgent consistently outperformed the LLM-only and Bandit-only baselines in terms of fix accuracy and reduction of irrelevant code changes.

\begin{table*}[!t]
\caption{Experimental Results}
\begin{center}
\begin{tabular}{|l|l|l|l|l|}
\hline
\textbf{Configuration} & \textbf{Fix Accuracy (\%)} & \textbf{False Positives (\%)} & \textbf{Avg. Iterations to Converge} & \textbf{Explanation Quality (/5)} \\
\hline
Bandit Only (Detection) & -- & 18.91 & -- & -- \\
LLM Only (Raw) & 74.32 & 13.57 & \textbf{1} & 2.7 \\
LLM Only (Fine Tuned) & 79.72 & 12.16 & \textbf{1} & 4.2 \\
SecureFixAgent (Base Model) & 81.08 & 12.16 & 4 & 4.3 \\
SecureFixAgent (Fine-tuned) & \textbf{87.83} & \textbf{8.11} & 3 & \textbf{4.5} \\
\hline
\end{tabular}
\label{tab:experimental_results}
\end{center}
\end{table*}

When operating with raw, instruction-tuned models, DeepSeek Coder 1.3B \cite{deepseekcoder} exhibited moderate baseline repair capability, with the larger variant achieving higher raw accuracy but generally requiring more iterations to converge. Qwen2.5 Coder 7B \cite{qwen25} showed comparatively strong raw performance and explanation quality, while CodeGemma 7B \cite{codegemma} and CodeLlama 7B \cite{codellama} produced balanced results with occasional extraneous formatting changes that did not affect security correctness.

\begin{figure}[htbp]
\centerline{\includegraphics[width=\linewidth, height=2in]{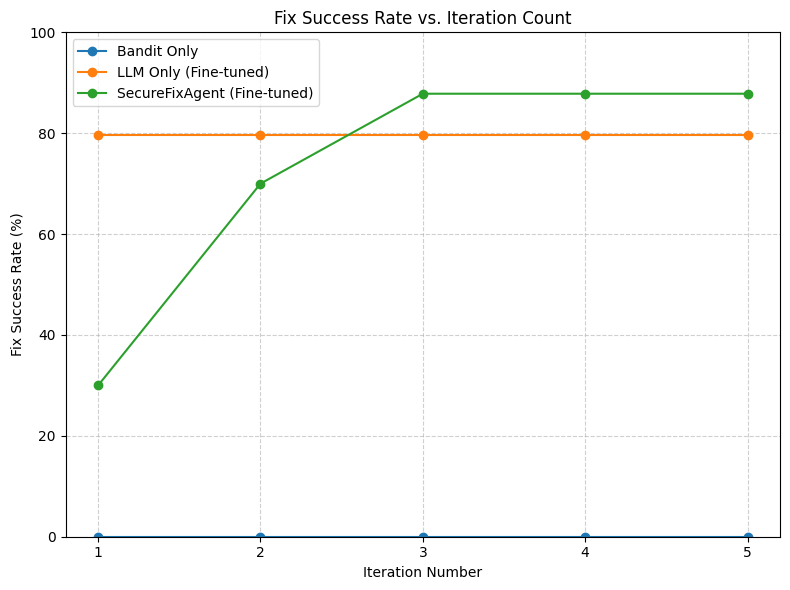}}
\caption{Fix success rate over iterations. Bandit-only remains at 0\%. LLM-only converges in the first iteration (79.72\%) with no further gains. SecureFixAgent improves iteratively, converging by the third iteration (87.83\%), demonstrating the value of detect--repair--validate cycles.}
\label{fig:fix_suc_vs_iter}
\end{figure}

Fine-tuning each model on the created vulnerability–repair dataset produced consistent gains. For example, DeepSeek Coder 6.7B \cite{deepseekcoder} improved in fix accuracy from 70.27\% (raw) to 77.02\% (fine-tuned) and reduced false positives from 14.86\% to 13.57\%. Qwen 2.5 Coder 7B \cite{qwen25} achieved the highest post-fine-tuning accuracy at 79.72\%, and its average iterations to convergence decreased to 3. CodeLlama 7B \cite{codellama} and CodeGemma 7B \cite{codegemma} likewise showed comparable improvements in accuracy and convergence (see Table \ref{tab:model_performance}).

\begin{table*}[!t]
\caption{LLM-only Performance Comparison}
\begin{center}
\begin{tabular}{|l|l|l|l|l|l|l|l|}
\hline
\textbf{Model} & \textbf{Params} & \textbf{Fix Acc. (Raw)} & \textbf{Fix Acc. (FT)} & \textbf{FP (\%) Raw} & \textbf{FP (\%) FT} & \textbf{Likert (Raw)} & \textbf{Likert (FT)} \\
\hline
DeepSeek Coder & 1.3B & 62.16 & 71.62 & 21.62 & 17.56 & 2.6 & 3.8 \\
DeepSeek Coder & 6.7B & 70.27 & 77.02 & 14.86 & 13.57 & 2.9 & 4.0 \\
Qwen 2.5 Coder & 3B & 72.97 & \textbf{79.72} & 18.91 & 14.86 & 2.9 & \textbf{4.3} \\
Qwen 2.5 Coder & 7B & \textbf{74.32} & \textbf{79.72} & \textbf{13.57} & \textbf{12.16} & 2.8 & 4.2 \\
CodeLlama & 7B & 60.81 & 67.56 & 14.86 & 14.86 & 2.8 & 3.8 \\
CodeGemma & 7B & 67.56 & 72.97 & 17.56 & 16.21 & \textbf{3.0} & 4.0 \\
\hline
\end{tabular}
\label{tab:model_performance}
\end{center}
\end{table*}

Critically, explanation clarity as perceived by human developers improved after fine-tuning. A mixed group of CS graduates, early-career employees, and experienced consultants (n = 15) rated clarity on a 1–5 Likert scale; average scores rose from $2.9/5$ (raw LLM) to $4.5/5$ (fine-tuned SecureFixAgent) for Qwen 7B \cite{qwen25}, with similar trends across other models (see Table \ref{tab:experimental_results}).

Across models, the full SecureFixAgent pipeline with fine-tuned LLMs consistently outperformed the LLM-only baseline, confirming the benefit of validation-driven iteration. The Bandit-only baseline, while high-precision at detection, naturally scored 0\% on fix accuracy since it does not perform repairs. These results validate the central design principle of SecureFixAgent: a hybrid, iterative detect--repair--validate loop increases patch correctness, reduces unnecessary edits, and produces explanations that developers find more comprehensible.

\begin{figure}[htbp]
\centerline{\includegraphics[width=\linewidth, height=2in]{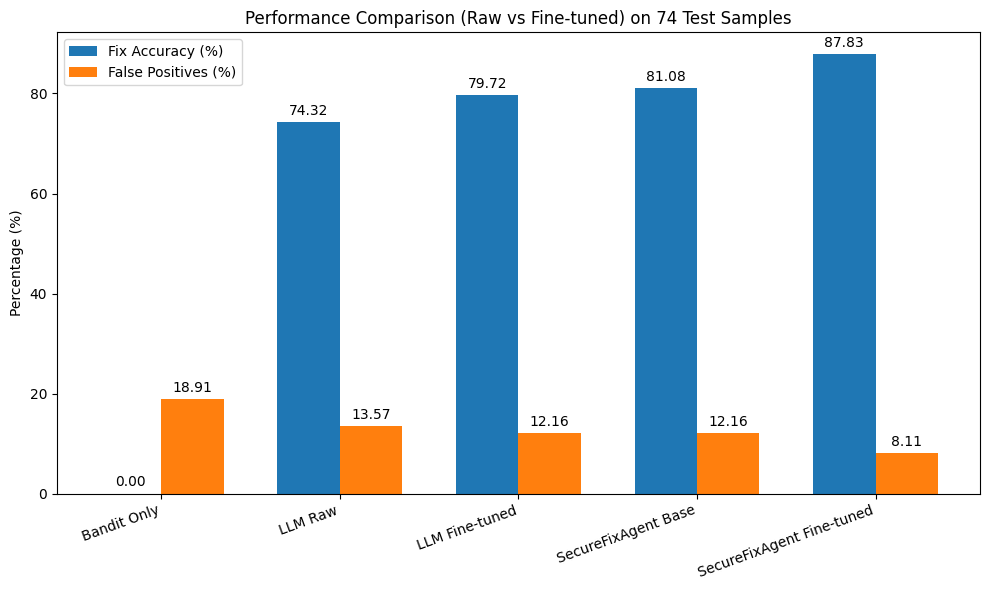}}
\caption{Fix accuracy and false positive rates. Bandit-only detects but cannot repair, with the highest false positives (18.91\%). LLM-only achieves moderate accuracy (74.32\%), improved with fine-tuning (79.72\%). SecureFixAgent yields the best results (87.83\% accuracy, 8.11\% false positives).}
\label{fig:perf_comparison}
\end{figure}

\section{Discussion and Future Work}
\label{dis_and_fut_work}

\subsection{Observed Strengths}
SecureFixAgent achieved 81–88\% higher fix accuracy than static analysis and 7–8\% over LLM-only baselines, with the greatest gains from fine-tuned models. The iterative detect--repair--validate loop reduced hallucinations, lowering false positives by up to 5.5\% compared to unvalidated LLM repair. In a 15-participant study, explanation clarity improved from 2.9 to 4.5, enhancing developer trust and usability.

\subsection{Limitations \& Future Work}
Despite its advantages, SecureFixAgent has some limitations. Its reliance on Bandit’s \cite{bandit} rule set restricts detection scope, leaving vulnerabilities outside predefined patterns unaddressed. Some patches—especially those involving distributed logic or multi-file edits—remained incomplete after the iteration limit. Robustness against adversarially crafted code also remains a concern, exposing potential gaps in detection and validation. These highlight the need to improve detection breadth, patch granularity, and adversarial resilience.

Future work will address these issues by integrating complementary analyzers (e.g., SonarQube \cite{sonarqube}, Semgrep \cite{semgrep}) to broaden coverage and reduce rule-set dependency, extending support to additional languages, and incorporating automated unit test generation with dynamic testing (fuzzing, runtime instrumentation). Together, these enhancements aim to improve coverage, robustness, explainability, and deployability of automated vulnerability remediation pipelines.

\section{Conclusion}
\label{concl}
We presented SecureFixAgent, a privacy-preserving framework that integrates static analysis with local, lightweight, code-specialized LLMs in an iterative detect--repair--validate loop for reliable and explainable vulnerability remediation. By combining Bandit’s rule-based precision with minimal LLM-generated patches and repeated validation, our approach improves fix accuracy, reduces false positives, and provides	 developer-trusted explanations on both synthetic and real-world datasets. Results confirm that fine-tuned, small-scale LLMs can address security-critical tasks without compromising performance or confidentiality. SecureFixAgent further enables integration into CI/CD pipelines and IDEs for continuous vulnerability mitigation. As LLMs evolve, future extensions may enhance repair sophistication, adaptability, and real-time assurance, reinforcing the value of hybrid static–LLM systems in secure software development.


\begin{thebibliography}{00}

\bibitem{gratton2025cybercrime}
P. Gratton, ``10 ways cybercrime impacts business,'' \textit{Investopedia}, Feb. 20, 2025. [Online]. Available: \url{https://www.investopedia.com/financial-edge/0112/3-ways-cyber-crime-impacts-business.aspx} (accessed: 01-Aug-2025).

\bibitem{he2023automating}
R. He \textit{et al.}, ``Automating dependency updates in practice: An exploratory study on GitHub Dependabot,'' \textit{IEEE Trans. Softw. Eng.}, vol. 49, no. 8, pp. 4004--4022, 2023.

\bibitem{chen2024deep}
Q. Chen \textit{et al.}, ``A deep dive into large language model code generation mistakes: What and why?,'' \textit{arXiv preprint arXiv:2411.01414}, 2024.

\bibitem{yao2024survey}
Y.~Yao \textit{et al.}, “A survey on large language model (LLM) security and privacy: The Good, The Bad, and The Ugly,” \textit{High-Confidence Comput.}, vol. 4, no. 2, p. 100211, 2024.

\bibitem{github2024security}
Kaaviya, “GitHub AI Copilot: Auto-Detect and Fix Code Vulnerabilities Instantly,” \textit{CyberPress}, Aug. 19, 2024. Available: https://cyberpress.org/auto-detect-and-fix-code-vulnerabilities/. [Online; accessed 02-Aug-2025].

\bibitem{harzevili2023}
N. S. Harzevili \textit{et al.}, ``Automatic static vulnerability detection for machine learning libraries: Are we there yet?,'' in \textit{Proc. IEEE Int. Symp. Softw. Rel. Eng. (ISSRE)}, 2023, pp. 795--806.

\bibitem{alashjaee2019}
A. M. Alashjaee \textit{et al.}, ``Dynamic taint analysis tools: A review,'' \textit{Int. J. Comput. Sci. Secur.}, vol. 13, no. 6, pp. 231--243, 2019.

\bibitem{codebert}
Z. Feng \textit{et al.}, ``CodeBERT: A pre-trained model for programming and natural languages,'' in \textit{Proc. EMNLP}, 2020.

\bibitem{juliet}
National Institute of Standards and Technology, ``Juliet test suite,'' 2023. [Online]. Available: \url{https://samate.nist.gov/SRD} (accessed: 01-Aug-2025).

\bibitem{sonarqube}
SonarSource, ``SonarQube: Continuous code quality,'' 2025. [Online]. Available: \url{https://www.sonarqube.org/} (accessed: 01-Aug-2025).

\bibitem{burpsuite}
PortSwigger, ``Burp Suite: Web vulnerability scanner,'' 2025. [Online]. Available: \url{https://portswigger.net/burp} (accessed: 01-Aug-2025).

\bibitem{zap}
OWASP, ``OWASP ZAP: Zed attack proxy,'' 2025. [Online]. Available: \url{https://owasp.org/www-project-zap/} (accessed: 01-Aug-2025).

\bibitem{bandit}
PyCQA, ``Bandit (1.8.6) [Computer software],'' 2025. [Online]. Available: \url{https://github.com/PyCQA/bandit} (accessed: 02-Aug-2025).

\bibitem{pylint}
Pylint contributors, ``Pylint,'' 2020. [Online]. Available: \url{https://pylint.readthedocs.io/en/latest/} (accessed: 02-Aug-2025).

\bibitem{semgrep}
Semgrep, Inc., ``Semgrep,'' 2025. [Online]. Available: \url{https://semgrep.dev/} (accessed: 02-Aug-2025).

\bibitem{draper}
R. Russell \textit{et al.}, ``Automated vulnerability detection in source code using deep representation learning,'' in \textit{Proc. 17th IEEE Int. Conf. Mach. Learn. Appl. (ICMLA)}, 2018, pp. 757--762.

\bibitem{vuldeepecker}
Z. Li \textit{et al.}, ``VulDeePecker: A deep learning-based system for vulnerability detection,'' in \textit{Proc. NDSS}, 2018.

\bibitem{zhou2025large}
X. Zhou \textit{et al.}, ``Large language model for vulnerability detection and repair: Literature review and the road ahead,'' \textit{ACM Trans. Softw. Eng. Methodol.}, vol. 34, no. 5, p. 145, 2025.

\bibitem{wang2025llm}
X. Wang \textit{et al.}, ``Practically implementing an LLM-supported collaborative vulnerability remediation process: A team-based approach,'' \textit{Comput. Secur.}, vol. 148, p. 104113, 2025.

\bibitem{repairnator}
M. Monperrus \textit{et al.}, ``Repairnator patches programs automatically,'' \textit{Ubiquity}, vol. 2019, no. July, p. 2, 2019.

\bibitem{t5apr}
R. Gharibi \textit{et al.}, ``T5APR: Empowering automated program repair across languages through checkpoint ensemble,'' \textit{J. Syst. Softw.}, vol. 214, p. 112083, 2024.

\bibitem{polycoder}
F. F. Xu \textit{et al.}, ``A systematic evaluation of large language models of code,'' in \textit{Proc. 6th ACM SIGPLAN Int. Symp. Mach. Program.}, 2022, pp. 1--10.

\bibitem{codet5}
Y. Wang \textit{et al.}, ``CodeT5: Identifier-aware unified pre-trained encoder-decoder models for code understanding and generation,'' in \textit{Proc. EMNLP}, 2021.

\bibitem{malbert}
H. Raff \textit{et al.}, ``MalBERT: Transformer-based malware detection,'' \textit{arXiv preprint arXiv:2109.09684}, 2021.

\bibitem{vulberta}
H. Hanif \textit{et al.}, ``Vulberta: Simplified source code pre-training for vulnerability detection,'' in \textit{Proc. Int. Joint Conf. Neural Netw. (IJCNN)}, 2022, pp. 1--8.

\bibitem{qwen25}
B. Hui \textit{et al.}, ``Qwen2.5-coder technical report,'' \textit{arXiv preprint arXiv:2409.12186}, 2024.

\bibitem{githubcopilot}
GitHub, Inc., “GitHub Copilot,” 2025. Available: https://github.com/features/copilot. [Online; accessed 03-Aug-2025].

\bibitem{mlx}
A. Hannun \textit{et al.}, ``MLX: Efficient and flexible machine learning on Apple silicon,'' 2023. [Online]. Available: \url{https://github.com/ml-explore} (accessed: 01-Aug-2025).

\bibitem{lora}
E. J. Hu \textit{et al.}, ``LoRA: Low-rank adaptation of large language models,'' \textit{arXiv preprint arXiv:2106.09685}, 2022.

\bibitem{malcodeai}
J. Gajjar \textit{et al.}, ``MalCodeAI: Autonomous vulnerability detection and remediation via language agnostic code reasoning,'' \textit{arXiv preprint arXiv:2507.10898}, 2025.

\bibitem{codegemma}
C. Team \textit{et al.}, ``Codegemma: Open code models based on Gemma,'' \textit{arXiv preprint arXiv:2406.11409}, 2024.

\bibitem{deepseekcoder}
Q. Zhu \textit{et al.}, ``DeepSeek-Coder-v2: Breaking the barrier of closed-source models in code intelligence,'' \textit{arXiv preprint arXiv:2406.11931}, 2024.

\bibitem{codellama}
B. Roziere \textit{et al.}, ``Code Llama: Open foundation models for code,'' \textit{arXiv preprint arXiv:2308.12950}, 2023.

\bibitem{cvebench}
Y. Zhu \textit{et al.}, ``CVE-Bench: A benchmark for AI agents’ ability to exploit real-world web application vulnerabilities,'' \textit{arXiv preprint arXiv:2503.17332}, 2025.

\bibitem{sun2023exploring}
S. Sun \textit{et al.}, ``Exploring security commits in Python,'' in \textit{Proc. 39th IEEE Int. Conf. Softw. Maint. Evol. (ICSME)}, 2023.

\bibitem{siddiq2022seceval}
M. L. Siddiq \textit{et al.}, ``SecurityEval dataset: Mining vulnerability examples to evaluate machine learning-based code generation techniques,'' in \textit{Proc. 1st Int. Workshop Mining Softw. Repositories Appl. Privacy Secur. (MSR4P\&S)}, 2022.

\end{thebibliography}
\end{document}